\documentclass[aps,prl,a4paper,superscriptaddress,twocolumn,showpacs,amsmath,amssymb,longbibliography]{revtex4} 

\usepackage{pdfpages}
\usepackage{graphicx,epsfig}

\usepackage[english]{babel}

\allowdisplaybreaks
\begin{document}

\newcommand{\cau}{\underline{c}^{\phantom{\dagger}}}
\newcommand{\ccu}{\underline{c}^\dagger}

\newcommand{\hu}{\underline{\hat{H}}}

\newcommand{\Dp}{\hat{\Delta}_{\text{p}}}
\newcommand{\Dh}{\hat{\Delta}_{\text{h}}}
\newcommand{\Deltap}{[\hat{\Delta}_{\text{p}}]}
\newcommand{\Deltah}{[\hat{\Delta}_{\text{h}}]}
\newcommand{\R}{\mathcal{R}}
\newcommand{\Rh}{\hat{\mathcal{R}}}

\newcommand{\uA}{|\underline{A},Ri\rangle}
\newcommand{\uB}{|\underline{B},R'j\rangle}

\newcommand{\E}{\mathcal{E}}
\newcommand{\G}{\mathcal{G}}
\newcommand{\Lag}{\mathcal{L}}
\newcommand{\M}{\mathcal{M}}
\newcommand{\N}{\mathcal{N}}
\newcommand{\U}{\mathcal{U}}
\newcommand{\F}{\mathcal{F}}
\newcommand{\V}{\mathcal{V}}
\newcommand{\C}{\mathcal{C}}
\newcommand{\I}{\mathcal{I}}
\newcommand{\s}{\sigma}
\newcommand{\up}{\uparrow}
\newcommand{\dw}{\downarrow}
\newcommand{\h}{\hat{H}}
\newcommand{\himp}{\hat{H}_{\text{imp}}}
\newcommand{\g}{\mathcal{G}^{-1}_0}
\newcommand{\D}{\mathcal{D}}
\newcommand{\A}{\mathcal{A}}
\newcommand{\projs}{\hat{\mathcal{S}}_d}
\newcommand{\proj}{\hat{\mathcal{P}}}
\newcommand{\K}{\textbf{k}}
\newcommand{\Q}{\textbf{q}}
\newcommand{\T}{\tau_{\ast}}
\newcommand{\io}{i\omega_n}
\newcommand{\eps}{\varepsilon}
\newcommand{\+}{\dag}
\newcommand{\su}{\uparrow}
\newcommand{\giu}{\downarrow}
\newcommand{\0}[1]{\textbf{#1}}
\newcommand{\ca}{c^{\phantom{\dagger}}}
\newcommand{\cc}{c^\dagger}
\newcommand{\aaa}{a^{\phantom{\dagger}}}
\newcommand{\aac}{a^\dagger}
\newcommand{\bba}{b^{\phantom{\dagger}}}
\newcommand{\bbc}{b^\dagger}
\newcommand{\da}{d^{\phantom{\dagger}}}
\newcommand{\dc}{d^\dagger}
\newcommand{\fa}{f^{\phantom{\dagger}}}
\newcommand{\fc}{f^\dagger}
\newcommand{\ha}{h^{\phantom{\dagger}}}
\newcommand{\hc}{h^\dagger}
\newcommand{\be}{\begin{equation}}
\newcommand{\ee}{\end{equation}}
\newcommand{\bea}{\begin{eqnarray}}
\newcommand{\eea}{\end{eqnarray}}
\newcommand{\ba}{\begin{eqnarray*}}
\newcommand{\ea}{\end{eqnarray*}}
\newcommand{\dagga}{{\phantom{\dagger}}}
\newcommand{\bR}{\mathbf{R}}
\newcommand{\bQ}{\mathbf{Q}}
\newcommand{\bq}{\mathbf{q}}
\newcommand{\bqp}{\mathbf{q'}}
\newcommand{\bk}{\mathbf{k}}
\newcommand{\bh}{\mathbf{h}}
\newcommand{\bkp}{\mathbf{k'}}
\newcommand{\bp}{\mathbf{p}}
\newcommand{\bL}{\mathbf{L}}
\newcommand{\bRp}{\mathbf{R'}}
\newcommand{\bx}{\mathbf{x}}
\newcommand{\by}{\mathbf{y}}
\newcommand{\bz}{\mathbf{z}}
\newcommand{\br}{\mathbf{r}}
\newcommand{\Ima}{{\Im m}}
\newcommand{\Rea}{{\Re e}}
\newcommand{\Pj}[2]{|#1\rangle\langle #2|}
\newcommand{\ket}[1]{\vert#1\rangle}
\newcommand{\bra}[1]{\langle#1\vert}
\newcommand{\setof}[1]{\left\{#1\right\}}
\newcommand{\fract}[2]{\frac{\displaystyle #1}{\displaystyle #2}}
\newcommand{\Av}[2]{\langle #1|\,#2\,|#1\rangle}
\newcommand{\av}[1]{\langle #1 \rangle}
\newcommand{\Mel}[3]{\langle #1|#2\,|#3\rangle}
\newcommand{\Avs}[1]{\langle \,#1\,\rangle_0}
\newcommand{\eqn}[1]{(\ref{#1})}
\newcommand{\Tr}{\mathrm{Tr}}

\newcommand{\Vb}{\bar{\mathcal{V}}}
\newcommand{\Vd}{\Delta\mathcal{V}}
\def\P{P_{02}}
\newcommand{\Pb}{\bar{P}_{02}}
\newcommand{\Pd}{\Delta P_{02}}
\def\t{\theta_{02}}
\newcommand{\tb}{\bar{\theta}_{02}}
\newcommand{\td}{\Delta \theta_{02}}
\newcommand{\Rb}{\bar{R}}
\newcommand{\Rd}{\Delta R}

\title{Gutzwiller Renormalization Group}

\author{Nicola Lanat\`a}
\affiliation{Department of Physics and Astronomy, Rutgers University, Piscataway, New Jersey 08856-8019, USA} 
\author{Yong-Xin Yao}
\affiliation{Ames Laboratory-U.S. DOE and Department of Physics and Astronomy, Iowa State 
University, Ames, Iowa IA 50011, USA}
\author{Xiaoyu Deng}
\affiliation{Department of Physics and Astronomy, Rutgers University, Piscataway, New Jersey 08856-8019, USA}
\author{Cai-Zhuang Wang}
\affiliation{Ames Laboratory-U.S. DOE and Department of Physics and Astronomy, Iowa State
University, Ames, Iowa IA 50011, USA}
\author{Kai-Ming~Ho}
\affiliation{Ames Laboratory-U.S. DOE and Department of Physics and Astronomy, Iowa State
University, Ames, Iowa IA 50011, USA}
\author{Gabriel Kotliar}
\affiliation{Department of Physics and Astronomy, Rutgers University, Piscataway, New Jersey 08856-8019, USA} 
\date{\today} 
%

\begin{abstract}

  We develop a variational scheme called ``Gutzwiller renormalization group'' (GRG),
  which enables us to calculate the ground state of Anderson impurity models (AIM)
  with arbitrary numerical precision.
  Our method can exploit the low-entanglement property of the ground state in
  combination with the framework of the Gutzwiller wavefunction, and suggests
  that the ground state of the AIM has a very simple structure, which
  can be represented very accurately in terms of a surprisingly small number
  of variational parameters.
  We perform benchmark calculations of the single-band AIM
  that validate our theory and indicate that the GRG
  might enable us to study complex systems beyond the reach of the other
  methods presently available and pave the way to interesting generalizations, e.g.,
  to nonequilibrium transport in nanostructures.

\end{abstract}

\maketitle

\emph{Introduction.}---
Impurity models are ubiquitous in condensed matter theory.
Originally, the AIM was deviced to describe magnetic
impurities embedded in metallic hosts and
heavy-fermion compound~\cite{Anderson_IM,Hewson}.
However, the AIM can be applied to describe many other
physical systems, such as quantum dots~\cite{qdot_1,qdot_2,qdot_12} and
dissipative two-level systems~\cite{dissipative_2level_systems}.
The importance of impurity models in condensed matter
has grown further with the emergence of
dynamical mean-field theory (DMFT)~\cite{DMFT}
and its success in describing strongly
correlated materials~\cite{LDA+U+DMFT,Held-review-DMFT,CDMFT-Jarrell,dmft_book}.
In fact, within DMFT, solving the many-body lattice problem amounts to
solve recursively an auxiliary AIM.

Among the many methodologies developed to solve impurity models,
one of the most successful is the numerical renormalization
group (NRG)~\cite{Wilson-NRG-review,Hewson},
which was originally deviced by Wilson in order to study the Kondo
problem, and was based on the idea of ``scaling'',
previously introduced by Anderson.
The starting point of the NRG procedure consists in
representing the AIM as a one-dimensional semi-infinite
tight-binding linear chain with the
impurity situated at one of the boundaries.
Within this representation, and by
using a logarithmic discretization of the density of states corresponding
to exponentially decaying hoppings (Wilson chains),
the NRG algorithm consists in a truncation scheme that enables to take
into account the coupling between the different length scales progressively,
starting from near the impurity (high energies) to longer distances
(low energies), while retaining a relatively small number of states
at each stage.

More recently, the NRG algorithm has been reinterpreted as a variational
method within the framework of the
``matrix product states''~\cite{PRB-DMRGvsNRG,Vaestrate_NRG-DMRG-MPS}.
From this perspective, the accuracy of NRG can be traced back into the
fact that, due to the locality of the interactions within the 
above mentioned one-dimensional representation of the AIM,
the ground state has very low entanglement, and can be consequently
accurately represented by a matrix product state with a
small bond dimension.
Thanks to this interpretation it was also possible to uncover
a connection between NRG and the density matrix renormalization group (DMRG)
method~\cite{DMRG-original-White-PRL,DMRG-original-White-PRB},
which can also be formulated as a variational
approximation within the framework of the matrix
product states~\cite{Vaestrate_DMRG-MPS}.
Eventually, these ideas resulted in an entirely new class of methods named
``variational renormalization group'' (VRG)~\cite{Vaestrate_review},
with broader applications with respect to both NRG and DMRG, encompassing, e.g.,
systems with dimension higher than 1.

In this work we introduce a variational technique to solve the AIM
called ``Gutzwiller renormalization group'' (GRG),
that combines the Gutzwiller variational method~\cite{Gutzwiller1,Gutzwiller2,Gutzwiller3}
with some of the above mentioned ideas underlying NRG
and the other VRG methods.
Similarly to
the ordinary Gutzwiller theory,
the GRG consists in optimizing variationally a wavefunction
represented as a Slater determinant
extended over the entire system --- which is \emph{infinite} ---
multiplied by a \emph{local} operator
(named ``Gutzwiller projector''), which
modifies the corresponding electron configurations. 
However, in GRG the action of the Gutzwiller projector is not limited only to
the correlated impurity.
In fact, in analogy with the NRG numerical procedure, 
in GRG the longer length scales
are taken into account with arbitrary accuracy by extending progressively
the action of the Gutzwiller projector also to
a portion of the one-dimensional bath lying nearby the impurity.
Thus, the size of the ``projected'' region is
increased until the desired level of precision is reached.

We perform numerical calculations that demonstrate the quality of the
GRG variational ansatz, and present arguments concerning
the scaling of the computational complexity of our algorithm
indicating that the GRG might enable
us to solve AIM beyond the reach of any other presently
available technique.
Another appealing property of our approach is that it enables us to
describe the ground state of the AIM in the thermodynamical
limit and at all length scales, while this is technically very difficult
with any other existing technique, including NRG, DMRG, and the
Bethe ansatz~\cite{Andrei1,Andrei2,Weigmann}.

\begin{figure}
\begin{center}
\includegraphics[width=8.3cm]{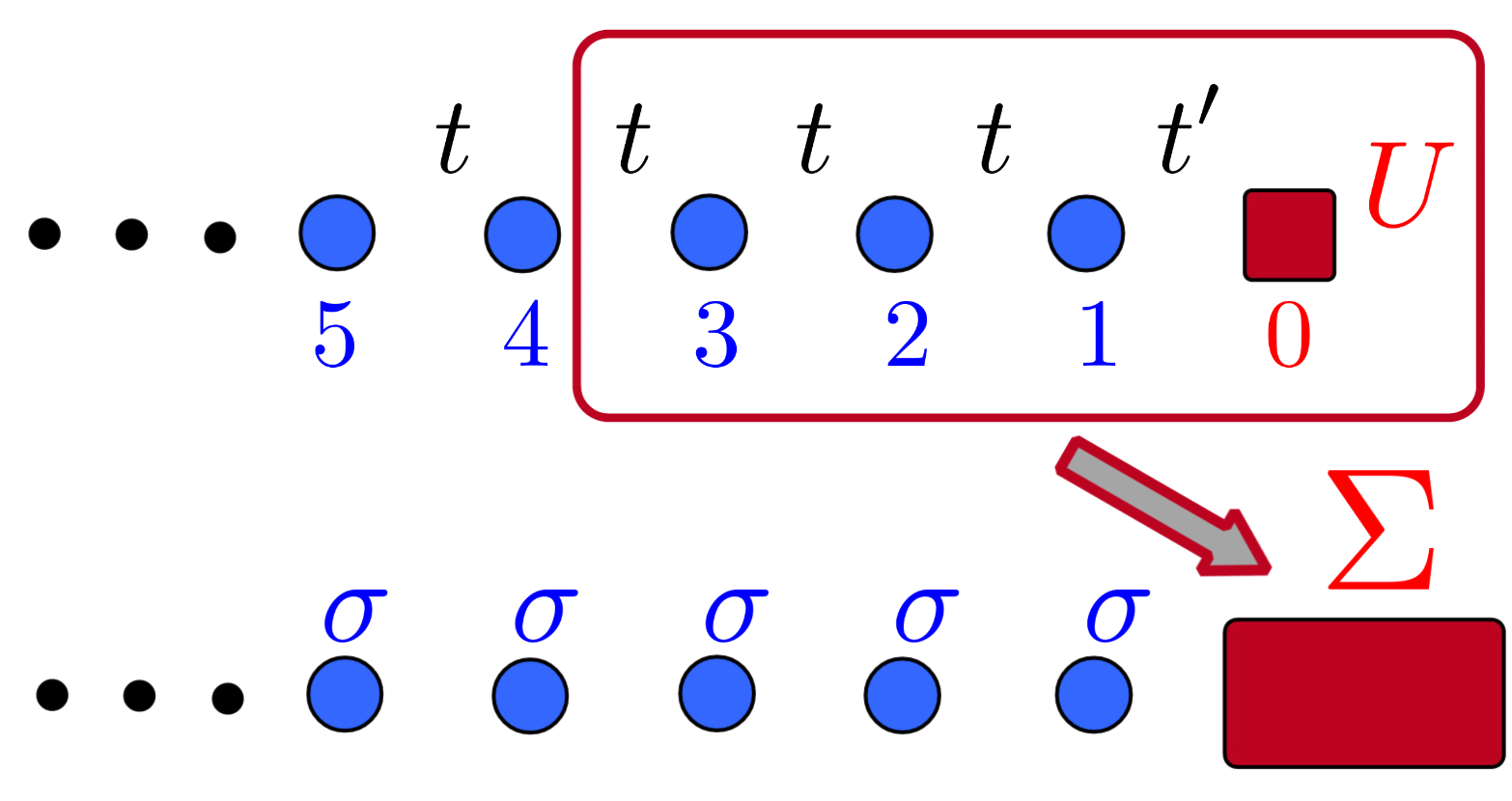}
\caption{Upper part. Representation of a single-band Anderson impurity model
  with first nearest neighbor hopping. Each site (bath and impurity) has
  the elementary spin degree of freedom $\sigma=\pm 1/2$. 
  Lower part. Equivalent representation of the same model
  where the impurity and a portion of the bath are formally considered
  as a larger impurity with effective elementary degree of freedom $\Sigma$.
}
\label{figure1}
\end{center}
\end{figure}

\emph{The Anderson Impurity Model}.---
For sake of simplicity, here our approach will be formulated for
the single-band AIM represented in
Fig.~\ref{figure1}, where the correlated
impurity is connected to a bath constituted by a linear chain
with first nearest-neighbor hopping:
\bea
\himp&=&\frac{U}{2}\left[1-\sum_\sigma\cc_{0,\sigma}\ca_{0,\sigma}\right]^2
\!\!-\!\!\sum_{\sigma=\pm \frac{1}{2}} t'\left(\cc_{1,\sigma}\ca_{0,\sigma}+\text{H.c.}\right)
\nonumber\\&-&
\sum_{\sigma=\pm \frac{1}{2}} \sum_{R=1}^\infty t\left(\cc_{R,\sigma}\ca_{R+1,\sigma}+\text{H.c.}\right)
\,.
\label{singleimp}
\eea
The generalization of our theory to generic impurity models is straightforward.

For later convenience, we point out
that the hybridization function of $\himp$
with respect to the correlated impurity is given by:
\be
\lim_{\eta\rightarrow 0^+} [\Delta(\omega+i\eta)]_{\sigma\sigma'}
= \delta_{\sigma\sigma'}\,g(\omega)\,,
\ee
where
\bea
g(\omega)\equiv
\left\{
\begin{array}{rl}
\Gamma\left[\frac{\omega}{D}-i\sqrt{1-\left(\frac{\omega}{D}\right)^2}\right] & \text{if}\;|\omega|\leq D \\
\Gamma\left[\frac{\omega}{D}-\text{sgn}(\omega)\sqrt{1-\left(\frac{\omega}{D}\right)^2}\right] & \text{if}\;|\omega|>D 
\,,~~~~
\end{array}
\right.
\label{g_Gamma}
\eea
$D=2t$, $\Gamma=t'^2/t$, and $\text{sgn}(\omega)$ is the sign function.

\emph{The Method}.---
In order to illustrate our method, it is convenient to observe that the
subsystem of $\himp$ consisting of the impurity and the first nearest $N$ bath sites
can be equivalently regarded as a larger impurity $\hat{S}_N[\{\aac_\Sigma,\aaa_{\Sigma}\}]$,
where the label $\Sigma = 1,..,2(N+1)$ 
runs over the sites $0,..,N$ and the corresponding spins $\sigma=\pm\frac{1}{2}$,
see Fig.~\ref{figure1}.
Within this definition, $\himp$ can be schematically represented as follows:
\bea
&&\himp = \hat{S}_N[\{\aac_\Sigma,\aaa_{\Sigma}\}]+\sum_{bb'}[B_N]_{bb'}\,\bbc_b\bba_{b'}
\nonumber\\&&\quad
+\sum_{b\Sigma}\left([V_N^\dagger]_{b\Sigma}\,\bbc_b\aaa_{\Sigma}+
[V_N^\dagga]_{\Sigma b}\,\aac_{\Sigma}\bba_b
\right)
\,,
\label{AIM}
\eea
where the hybridization function $\Delta_N$ of $\hat{S}_N$ 
\be
\Delta_N(z)\equiv
V^\dagga_N\,\frac{1}{z-B_N}\,V_N^\dagger
\label{Deltaqp}
\ee
is a well defined $M\times M$ matrix, where $M\equiv 2(N+1)$.

Our approach consists
in optimizing variationally a Gutzwiller wavefunction represented as follows:
\be
\ket{\Psi_N}\equiv\proj_N\,\ket{\Psi_0}\,,
\label{GW}
\ee
where $\ket{\Psi_0}$ is a generic Slater determinant, and
$\proj_N$, which is called ``Gutzwiller projector'',
is the most general operator acting within the subsystem $\hat{S}_N$.
As in the ordinary Gutzwiller approximation scheme,
in order to simplify the task of calculating the expectation value of
the Hamiltonian [Eq.~\eqref{AIM}], the
variational freedom is further restricted by assuming the following conditions:
\bea
\Av{\Psi_0}{\proj_N^\dagger\proj^\dagga_N} \!\!&=&\!\! \langle \Psi_0 |\Psi_0 \rangle = 1
\label{gc1}\\
\Av{\Psi_0}{\proj_N^\dagger\proj^\dagga_N\,\aac_\Sigma\aaa_{\Sigma'}} \!\!&=&\!\!
\Av{\Psi_0}{\aac_\Sigma\aaa_{\Sigma'}}\;\,\forall\;\Sigma,\Sigma'
\label{gc2}
\,,
\eea
which are called ``Gutzwiller constraints''.
We point out that, while for bulk systems the Gutzwiller variational
ansatz is supplemented by the so called
``Gutzwiller approximation''~\cite{Gutzwiller3},
for impurity models the method is purely variational,
and no further approximation is needed~\cite{Our-PRX,my-transport-1,my-transport-2}.

For $N=0$ the procedure described above
reduces to applying the Gutzwiller
projector only onto the correlated impurity --- which is the
standard Gutzwiller approach for the AIM used, e.g., in
Refs.~\cite{my-transport-1,my-transport-2}.
In this work, instead,
the region of action of the Gutzwiller projector is extended
systematically by
increasing $N$ until the desired level of accuracy is obtained.

The most evident difficulty to be overcome is that the number of complex
independent parameters defining $\proj_N$ scales as $2^{2M}$.
In fact, this makes the task of minimizing the total energy with respect to
the wavefunction [Eq.~\eqref{GW}]
nontrivial already for small $N$.
However, fortunately, this technical problem
can be efficiently solved even for relatively large $N$
thanks to the method of Refs.~\cite{nostro,Our-PRX,Our-PRB-SB},
which is summarized in the supplemental material for completeness.
A remarkable aspect of this numerical scheme is that it enables us
to map the above nonlinear constrained minimization into
the much simpler task of calculating iteratively
the ground state $\ket{\Phi}$ of
a finite AIM represented as follows:
\bea
&&\h^{\text{emb}}[\D,\lambda^c]
\equiv\hat{S}_N[\{{a}^\dagger_{\Sigma},{a}^\dagga_{\Sigma}\}]
\nonumber\\&&\quad
+\sum_{s\Sigma=1}^M \left(
\D_{s\Sigma}\,
{a}^\dagger_{\Sigma}{f}^\dagga_{s}+\text{H.c.}\right)
+\sum_{ss'=1}^M \lambda^c_{ss'}
{f}^\dagga_{s'}{f}^\dagger_{s}
\label{h-embimp}
\eea
at half-filling, i.e., within the subspace such that:
\be
\left[
\sum_{\Sigma=1}^M{a}^\dagger_{\Sigma}{a}^\dagga_{\Sigma}
+ \sum_{s=1}^M{f}^\dagger_{s}{f}^\dagga_{s}
\right]\ket{\Phi} = M\,\ket{\Phi}\,.
\ee
The complex coefficients $\D_{s\Sigma}$ and $\lambda^c_{ss'}$ have
to be determined numerically following the procedure summarized
in the supplemental material.
The number of AIM represented as in Eq.~\eqref{h-embimp} to be solved in order to
converge scales as $o(M^2)$, and they can be solved
independently (in parallel) if necessary.

As shown in the supplemental material,
the ground state $\ket{\Phi}$ of $\h^{\text{emb}}$ 
for the converged parameters $\D$ and $\lambda^c$
encodes the expectation value
with respect to the corresponding Gutzwiller wavefunction
Eq.~\eqref{GW} of any observable $\hat{A}$ in $\hat{S}_N$, i.e.:
\be
\Av{\Psi_N}{\hat{A}[\{{a}^\dagger_{\Sigma},{a}^\dagga_{\Sigma}\}]}=
\Av{\Phi}{\hat{A}[\{{a}^\dagger_{\Sigma},{a}^\dagga_{\Sigma}\}]}\,.
\label{phi-imp}
\ee
For this reason, as in Ref.~\cite{Our-PRX}, here we call $\h^{\text{emb}}$
``embedding Hamiltonian''.
Concerning the fact that in Eq.~\eqref{h-embimp}
the subsystem $\hat{S}_N$ is coupled with a bath of equal dimension,
it is interesting to observe that, at least in principle,
a Hamiltonian whose bath has the same dimension of the
impurity is sufficient to represent \emph{exactly} the impurity
ground-state properties of any system (no matter how big it is).
This fact can be readily
demonstrated making use of the Schmidt decomposition~\cite{Entanglement-review,DMET}.

Let us now discuss how the computational effort to calculate
the ground state of $\h^{\text{emb}}$ scales with $M$.
In order to answer this question it is important to note
that $\h^{\text{emb}}$ is \emph{quadratic} for all degrees of freedom made exception for 
those corresponding to 
$\{\cc_{0,\sigma},\ca_{0,\sigma}\}$ within the original representation of the
AIM, see Eq.~\eqref{singleimp}.
Thanks to this observation, Eq.~\eqref{h-embimp} can be always
transformed into a finite one-dimensional chain
by tridiagonalization~\cite{Hewson}.
This enables us to calculate $\ket{\Phi}$ using efficient
techniques such as DMRG~\cite{PRB-DMRGvsNRG,Vaestrate_NRG-DMRG-MPS},
whose computational cost grows only polynomially with $M$
rather than exponentially.
Note that in order to perform the specific calculations that we
are going to discuss in this work it has not been actually
necessary to resort to this stratagem.
However, this might be needed in order to study impurity models
more complicated than Eq.~\eqref{singleimp}.

We point out that the possibility to incorporate DMRG within
our algorithm constitutes a further connection between GRG and
the VRG methods mentioned in the introduction
(see also Refs.~\cite{MPPS-RC,MPPS-PRB}),
as it reflects the fact that our approach can enforce
the low-entanglement property of the ground state
within the framework of the Gutzwiller wavefunction,
and exploit simultaneously both of these
ideas~\footnote{Note that within the framework of
Refs.~\cite{MPPS-RC,MPPS-PRB}
the variational Monte Carlo method was used in
order to optimize the state, as it was
not technically possible to apply the more efficient
DMRG algorithm.}.

From the considerations above we deduce that the computational
complexity of the GRG algorithm scales polynomially with $N$, i.e.,
with the size of the region of action of the Gutzwiller projector.
Our remaining task is to understand how big $N$ has to be
in order to describe accurately the ground state of the AIM.

\emph{Benchmark calculations}.---
In order to assess the quality of the GRG variational ansatz,
here we perform benchmark calculations of the single-band AIM
at half-filling, see Eq.~\eqref{singleimp}.

\begin{figure}
\begin{center}
\includegraphics[width=8.8cm]{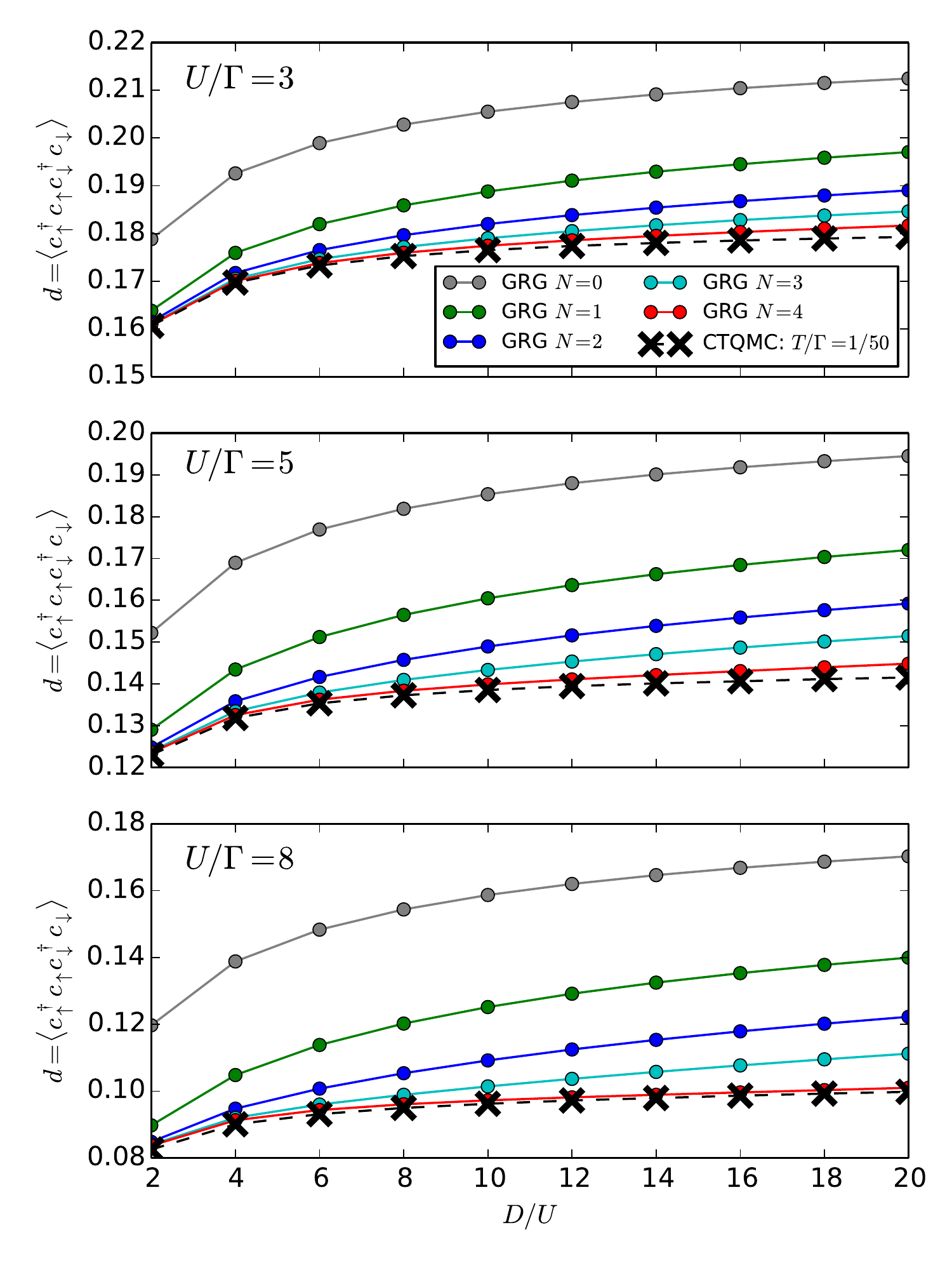}
\caption{(Color online) Convergence of the GRG
expectation value of the double occupancy $d$
with respect to the number of conduction-bath sites $N$
included within the projected region.
The result is shown for different $U/\Gamma$ and $D/U$
in comparison with CTQMC (black crosses).
}
\label{figure2}
\end{center}
\end{figure}

In Fig.~\ref{figure2} is shown the behavior of the 
impurity double occupancy $d$
as a function of $D/U$ for different values of $U/\Gamma$.
The GRG results obtained with different values of $N$
are shown in comparison with continuous time quantum Monte
Carlo (CTQMC)~\cite{ctqmc_hybr-exp_Rubtsov,ctqmc}
as implemented in TRIQS~\cite{TRIQS}, which
were obtained at temperature $T/\Gamma=0.02$.
From these calculations it emerges that 
the $N=4$ GRG wavefunction is already sufficient to
reproduce very accurately the CTQMC double occupancy
for all of the interaction parameters considered.
However, we observe
that the convergence of $d$ with respect to $N$
becomes increasingly slower for larger $D/U$,
while it is not very sensitive to $U/\Gamma$.
It is interesting to compare this trend
with the behavior of the dimension $\xi_K$ of the Kondo cloud.
From the Bethe ansatz solution of the single band AIM
we know that, in the large-$D$ limit, $\xi_K$ is given by~\cite{Hewson}:
\be
\xi_K\equiv v_F/T_K \sim
\frac{D}{U}\sqrt{\frac{2U}{\Gamma}}
e^{\frac{\pi}{8}\frac{U}{\Gamma}+\frac{\pi}{2}\frac{\Gamma}{U}}
\label{xiK}
\ee
(in units of the lattice spacing).
As we see from Eq.~\eqref{xiK}, $\xi_K$ 
diverges exponentially with $U/\Gamma$.
Furthermore, we note that $\xi_K$ is much bigger than the value of $N$
necessary to achieve convergence within the GRG.
In fact, for instance, substituting the values $D/U=10$ and $U/\Gamma=5$
in Eq.~\eqref{xiK} we find that $\xi_K\sim 275 \gg 5$.
Thus, we deduce that these two length scales are not
directly related, as one might naively expect.

\begin{figure}
\begin{center}
\includegraphics[width=8.8cm]{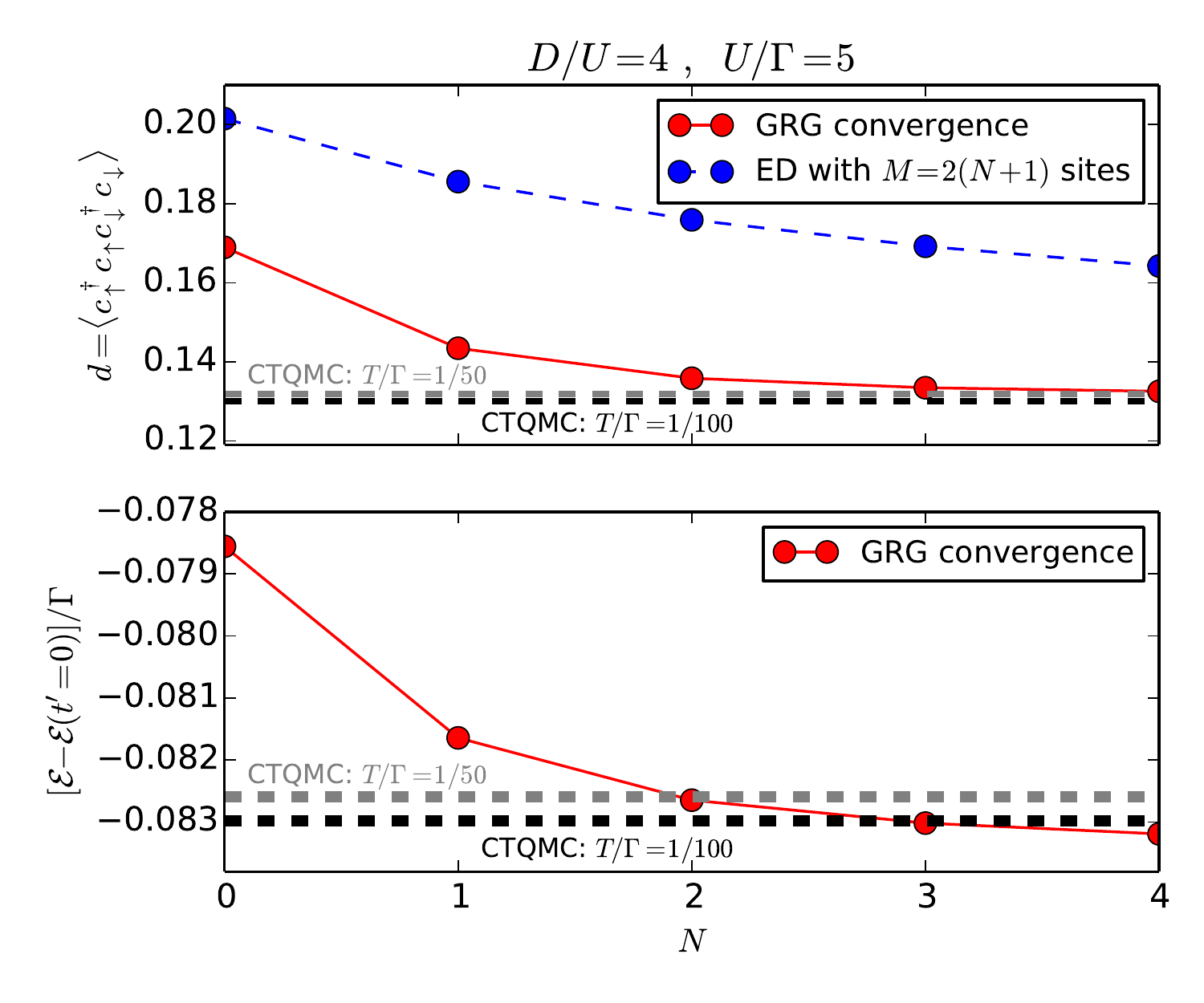}
\caption{(Color online) Upper panel: Convergence as a function of $N$
  of the GRG double occupancy $d$ for $D/U = 4$ and $U/\Gamma = 3$
  in comparison with exact diagonalization for the truncated system
  composed by $2(N+1)$ sites.
  Lower panel: Corresponding convergence of the total energy $\mathcal{E}$.
  The results calculated with CTQMC are indicated by the
  dotted horizontal lines.
}
\label{figure3}
\end{center}
\end{figure}

In Fig.~\ref{figure3} is shown the behavior of the
impurity double occupancy
and of the total energy 
calculated with GRG 
as a function of the convergence parameter $N$.
Note that the zero of the total energy 
is conventionally assumed to be the ground state energy
of Eq.~\eqref{singleimp} with $t'=0$.
These data are shown in comparison with CTQMC and with
the exact diagonalization results obtained by
taking into account only the first $2(N+1)$ sites of
Eq.~\eqref{singleimp} (and discarding the others).

We observe that both the GRG energy and double occupancy converge 
very rapidly to their respective exact values.
From this observation we argue that the above mentioned GRG convergence
concerns the entire ground state of the AIM,
and not only the expectation value of the impurity degrees of freedom.

From the comparison between GRG and exact diagonalization it emerges
that the dimension of the auxiliary AIM [Eq.~\eqref{h-embimp}]
to be solved in order to calculate the GRG solution of any order $N$
is much smaller than
the dimension of the truncated AIM that would enable us 
to solve the problem with comparable precision
directly with DMRG~\cite{Vaestrate_NRG-DMRG-MPS}.
Thus, as discussed before, using DMRG
to solve iteratively Eq.~\eqref{h-embimp} within the GRG algorithm
seems to be a more convenient option.

\emph{Conclusions.}---
In summary, we have developed a variational method called GRG,
that takes advantage simultaneously of two among the most
powerful ideas in condensed matter theory:
(1) the Gutzwiller wavefunction, that enables us to incorporate the
most general single-particle wavefunction (Slater determinant)
within the variational space, thus reducing the many-body
problem to correcting variationally the corresponding electron
configurations with the ``Gutzwiller projector'';
(2) the VRG methods, that enable us to exploit the fact
that the ground state of the AIM has
low-entanglement (once the problem is formulated in
one dimension).
Using the GRG, we have shown that the ground state of the AIM has
a very simple structure, which can be represented very
accurately in terms of a surprisingly
small number of variational parameters.
These insights resulted in an efficient algorithm that
might enable us to study complex systems beyond the reach of any other
method presently available.
Another remarkable property of our approach is that it
enables us to describe the ground state of the AIM
directly in the thermodynamical limit and at all length scales,
while this is technically very difficult with the other
VRG methods, which require to truncate part of the chain.
Our work paves the way to several generalizations.
In particular, it will be very interesting to generalize it
to finite temperatures~\cite{Our-Temperature} and
to nonequilibrium transport in nanostructures, see
Refs.~\cite{my-transport-1,my-transport-2}.
In fact, the physics of the AIM out of equilibrium is not still well
understood, and different methods seem to produce different results even for
simple systems such as the single band AIM~\cite{Dirks-comparative-noneq_EPL}.

\begin{acknowledgments}
  We thank Natan Andrei and Michele Fabrizio for useful discussions.
  N.L., X.D. and G.K. were supported by U.S. DOE Office of
  Basic Energy Sciences under Grant No. DE-FG02-99ER45761.
  Research at Ames Laboratory supported by the U.S.
  Department of Energy, Office of Basic Energy Sciences,
  Division of Materials Sciences and Engineering. 
  Ames Laboratory is operated for the U.S. Department of Energy
  by Iowa State University under Contract No. DE-AC02-07CH11358.
\end{acknowledgments}


\begin{thebibliography}{38}
\expandafter\ifx\csname natexlab\endcsname\relax\def\natexlab#1{#1}\fi
\expandafter\ifx\csname bibnamefont\endcsname\relax
  \def\bibnamefont#1{#1}\fi
\expandafter\ifx\csname bibfnamefont\endcsname\relax
  \def\bibfnamefont#1{#1}\fi
\expandafter\ifx\csname citenamefont\endcsname\relax
  \def\citenamefont#1{#1}\fi
\expandafter\ifx\csname url\endcsname\relax
  \def\url#1{\texttt{#1}}\fi
\expandafter\ifx\csname urlprefix\endcsname\relax\def\urlprefix{URL }\fi
\providecommand{\bibinfo}[2]{#2}
\providecommand{\eprint}[2][]{\url{#2}}

\bibitem[{\citenamefont{Anderson}(1961)}]{Anderson_IM}
\bibinfo{author}{\bibfnamefont{P.~W.} \bibnamefont{Anderson}},
  \bibinfo{journal}{Phys. Rev.} \textbf{\bibinfo{volume}{124}},
  \bibinfo{pages}{41} (\bibinfo{year}{1961}).

\bibitem[{\citenamefont{Hewson}(1997)}]{Hewson}
\bibinfo{author}{\bibfnamefont{A.~C.} \bibnamefont{Hewson}},
  \emph{\bibinfo{title}{The $\text{Kondo}$ $\text{Problem}$ to $\text{Heavy}$
  $\text{Fermions}$}} (\bibinfo{publisher}{Cambridge University Press},
  \bibinfo{year}{1997}).

\bibitem[{\citenamefont{Madhavan et~al.}(1998)\citenamefont{Madhavan, Chen,
  Jamneala, Crommie, and Wingreen}}]{qdot_1}
\bibinfo{author}{\bibfnamefont{V.}~\bibnamefont{Madhavan}},
  \bibinfo{author}{\bibfnamefont{W.}~\bibnamefont{Chen}},
  \bibinfo{author}{\bibfnamefont{T.}~\bibnamefont{Jamneala}},
  \bibinfo{author}{\bibfnamefont{M.}~\bibnamefont{Crommie}}, \bibnamefont{and}
  \bibinfo{author}{\bibfnamefont{N.~S.} \bibnamefont{Wingreen}},
  \bibinfo{journal}{Science} \textbf{\bibinfo{volume}{280}},
  \bibinfo{pages}{567} (\bibinfo{year}{1998}).

\bibitem[{\citenamefont{Cronenwett et~al.}(1998)\citenamefont{Cronenwett,
  Oosterkamp, and Kouwenhoven}}]{qdot_2}
\bibinfo{author}{\bibfnamefont{S.~M.} \bibnamefont{Cronenwett}},
  \bibinfo{author}{\bibfnamefont{T.~H.} \bibnamefont{Oosterkamp}},
  \bibnamefont{and} \bibinfo{author}{\bibfnamefont{L.~P.}
  \bibnamefont{Kouwenhoven}}, \bibinfo{journal}{Science}
  \textbf{\bibinfo{volume}{281}}, \bibinfo{pages}{540} (\bibinfo{year}{1998}).

\bibitem[{\citenamefont{van~der Wiel et~al.}(2002)\citenamefont{van~der Wiel,
  De~Franceschi, Elzerman, Fujisawa, Tarucha, and Kouwenhoven}}]{qdot_12}
\bibinfo{author}{\bibfnamefont{W.~G.} \bibnamefont{van~der Wiel}},
  \bibinfo{author}{\bibfnamefont{S.}~\bibnamefont{De~Franceschi}},
  \bibinfo{author}{\bibfnamefont{J.~M.} \bibnamefont{Elzerman}},
  \bibinfo{author}{\bibfnamefont{T.}~\bibnamefont{Fujisawa}},
  \bibinfo{author}{\bibfnamefont{S.}~\bibnamefont{Tarucha}}, \bibnamefont{and}
  \bibinfo{author}{\bibfnamefont{L.~P.} \bibnamefont{Kouwenhoven}},
  \bibinfo{journal}{Rev. Mod. Phys.} \textbf{\bibinfo{volume}{75}},
  \bibinfo{pages}{1} (\bibinfo{year}{2002}).

\bibitem[{\citenamefont{Vlad\'ar and
  Zawadowski}(1983)}]{dissipative_2level_systems}
\bibinfo{author}{\bibfnamefont{K.}~\bibnamefont{Vlad\'ar}} \bibnamefont{and}
  \bibinfo{author}{\bibfnamefont{A.}~\bibnamefont{Zawadowski}},
  \bibinfo{journal}{Phys. Rev. B} \textbf{\bibinfo{volume}{28}},
  \bibinfo{pages}{1564} (\bibinfo{year}{1983}).

\bibitem[{\citenamefont{Georges et~al.}(1996)\citenamefont{Georges, Kotliar,
  Krauth, and Rozenberg}}]{DMFT}
\bibinfo{author}{\bibfnamefont{A.}~\bibnamefont{Georges}},
  \bibinfo{author}{\bibfnamefont{G.}~\bibnamefont{Kotliar}},
  \bibinfo{author}{\bibfnamefont{W.}~\bibnamefont{Krauth}}, \bibnamefont{and}
  \bibinfo{author}{\bibfnamefont{M.~J.} \bibnamefont{Rozenberg}},
  \bibinfo{journal}{Rev. Mod. Phys.} \textbf{\bibinfo{volume}{68}},
  \bibinfo{pages}{13} (\bibinfo{year}{1996}).

\bibitem[{\citenamefont{Kotliar et~al.}(2006)\citenamefont{Kotliar, Savrasov,
  Haule, Oudovenko, Parcollet, and Marianetti}}]{LDA+U+DMFT}
\bibinfo{author}{\bibfnamefont{G.}~\bibnamefont{Kotliar}},
  \bibinfo{author}{\bibfnamefont{S.~Y.} \bibnamefont{Savrasov}},
  \bibinfo{author}{\bibfnamefont{K.}~\bibnamefont{Haule}},
  \bibinfo{author}{\bibfnamefont{V.~S.} \bibnamefont{Oudovenko}},
  \bibinfo{author}{\bibfnamefont{O.}~\bibnamefont{Parcollet}},
  \bibnamefont{and} \bibinfo{author}{\bibfnamefont{C.~A.}
  \bibnamefont{Marianetti}}, \bibinfo{journal}{Rev. Mod. Phys.}
  \textbf{\bibinfo{volume}{78}}, \bibinfo{eid}{865} (\bibinfo{year}{2006}).

\bibitem[{\citenamefont{Held et~al.}(2006)\citenamefont{Held, Nekrasov, Keller,
  Eyert, Bl\"umer, McMahan, Scalettar, Pruschke, Anisimov, and
  Vollhardt}}]{Held-review-DMFT}
\bibinfo{author}{\bibfnamefont{K.}~\bibnamefont{Held}},
  \bibinfo{author}{\bibfnamefont{A.}~\bibnamefont{Nekrasov}},
  \bibinfo{author}{\bibfnamefont{G.}~\bibnamefont{Keller}},
  \bibinfo{author}{\bibfnamefont{V.}~\bibnamefont{Eyert}},
  \bibinfo{author}{\bibfnamefont{N.}~\bibnamefont{Bl\"umer}},
  \bibinfo{author}{\bibfnamefont{A.~K.} \bibnamefont{McMahan}},
  \bibinfo{author}{\bibfnamefont{R.~T.} \bibnamefont{Scalettar}},
  \bibinfo{author}{\bibfnamefont{T.}~\bibnamefont{Pruschke}},
  \bibinfo{author}{\bibfnamefont{V.~I.} \bibnamefont{Anisimov}},
  \bibnamefont{and}
  \bibinfo{author}{\bibfnamefont{D.}~\bibnamefont{Vollhardt}},
  \bibinfo{journal}{Phys. Stat. Sol. (B)} \textbf{\bibinfo{volume}{243}},
  \bibinfo{pages}{2599} (\bibinfo{year}{2006}).

\bibitem[{\citenamefont{Maier et~al.}(2005)\citenamefont{Maier, Jarrell,
  Pruschke, and Hettler}}]{CDMFT-Jarrell}
\bibinfo{author}{\bibfnamefont{T.}~\bibnamefont{Maier}},
  \bibinfo{author}{\bibfnamefont{M.}~\bibnamefont{Jarrell}},
  \bibinfo{author}{\bibfnamefont{T.}~\bibnamefont{Pruschke}}, \bibnamefont{and}
  \bibinfo{author}{\bibfnamefont{M.~H.} \bibnamefont{Hettler}},
  \bibinfo{journal}{Rev. Mod. Phys.} \textbf{\bibinfo{volume}{77}},
  \bibinfo{pages}{1027} (\bibinfo{year}{2005}).

\bibitem[{\citenamefont{Anisimov and Izyumov}(2010)}]{dmft_book}
\bibinfo{author}{\bibfnamefont{V.}~\bibnamefont{Anisimov}} \bibnamefont{and}
  \bibinfo{author}{\bibfnamefont{Y.}~\bibnamefont{Izyumov}},
  \emph{\bibinfo{title}{Electronic Structure of Strongly Correlated Materials}}
  (\bibinfo{publisher}{Springer}, \bibinfo{year}{2010}).

\bibitem[{\citenamefont{Wilson}(1975)}]{Wilson-NRG-review}
\bibinfo{author}{\bibfnamefont{K.~G.} \bibnamefont{Wilson}},
  \bibinfo{journal}{Rev. Mod. Phys.} \textbf{\bibinfo{volume}{47}},
  \bibinfo{pages}{773} (\bibinfo{year}{1975}).

\bibitem[{\citenamefont{Saberi et~al.}(2008)\citenamefont{Saberi, Weichselbaum,
  and von Delft}}]{PRB-DMRGvsNRG}
\bibinfo{author}{\bibfnamefont{H.}~\bibnamefont{Saberi}},
  \bibinfo{author}{\bibfnamefont{A.}~\bibnamefont{Weichselbaum}},
  \bibnamefont{and} \bibinfo{author}{\bibfnamefont{J.}~\bibnamefont{von
  Delft}}, \bibinfo{journal}{Phys. Rev. B} \textbf{\bibinfo{volume}{78}},
  \bibinfo{pages}{035124} (\bibinfo{year}{2008}).

\bibitem[{\citenamefont{Weichselbaum et~al.}(2009)\citenamefont{Weichselbaum,
  Verstraete, Schollw\"ock, Cirac, and von Delft}}]{Vaestrate_NRG-DMRG-MPS}
\bibinfo{author}{\bibfnamefont{A.}~\bibnamefont{Weichselbaum}},
  \bibinfo{author}{\bibfnamefont{F.}~\bibnamefont{Verstraete}},
  \bibinfo{author}{\bibfnamefont{U.}~\bibnamefont{Schollw\"ock}},
  \bibinfo{author}{\bibfnamefont{J.~I.} \bibnamefont{Cirac}}, \bibnamefont{and}
  \bibinfo{author}{\bibfnamefont{J.}~\bibnamefont{von Delft}},
  \bibinfo{journal}{Phys. Rev. B} \textbf{\bibinfo{volume}{80}},
  \bibinfo{pages}{165117} (\bibinfo{year}{2009}).

\bibitem[{\citenamefont{White}(1992)}]{DMRG-original-White-PRL}
\bibinfo{author}{\bibfnamefont{S.~R.} \bibnamefont{White}},
  \bibinfo{journal}{Phys. Rev. Lett.} \textbf{\bibinfo{volume}{69}},
  \bibinfo{pages}{2863} (\bibinfo{year}{1992}).

\bibitem[{\citenamefont{White}(1993)}]{DMRG-original-White-PRB}
\bibinfo{author}{\bibfnamefont{S.~R.} \bibnamefont{White}},
  \bibinfo{journal}{Phys. Rev. B} \textbf{\bibinfo{volume}{48}},
  \bibinfo{pages}{10345} (\bibinfo{year}{1993}).

\bibitem[{\citenamefont{Verstraete et~al.}(2004)\citenamefont{Verstraete,
  Porras, and Cirac}}]{Vaestrate_DMRG-MPS}
\bibinfo{author}{\bibfnamefont{F.}~\bibnamefont{Verstraete}},
  \bibinfo{author}{\bibfnamefont{D.}~\bibnamefont{Porras}}, \bibnamefont{and}
  \bibinfo{author}{\bibfnamefont{J.~I.} \bibnamefont{Cirac}},
  \bibinfo{journal}{Phys. Rev. Lett.} \textbf{\bibinfo{volume}{93}},
  \bibinfo{pages}{227205} (\bibinfo{year}{2004}).

\bibitem[{\citenamefont{Vaestrate et~al.}(2008)\citenamefont{Vaestrate, Cirac,
  and Murg}}]{Vaestrate_review}
\bibinfo{author}{\bibfnamefont{F.}~\bibnamefont{Vaestrate}},
  \bibinfo{author}{\bibfnamefont{J.}~\bibnamefont{Cirac}}, \bibnamefont{and}
  \bibinfo{author}{\bibfnamefont{V.}~\bibnamefont{Murg}},
  \bibinfo{journal}{Adv. Phys.} \textbf{\bibinfo{volume}{57}},
  \bibinfo{pages}{143} (\bibinfo{year}{2008}).

\bibitem[{\citenamefont{Gutzwiller}(1963)}]{Gutzwiller1}
\bibinfo{author}{\bibfnamefont{M.~C.} \bibnamefont{Gutzwiller}},
  \bibinfo{journal}{Phys. Rev. Lett.} \textbf{\bibinfo{volume}{10}},
  \bibinfo{pages}{159} (\bibinfo{year}{1963}).

\bibitem[{\citenamefont{Gutzwiller}(1964)}]{Gutzwiller2}
\bibinfo{author}{\bibfnamefont{M.~C.} \bibnamefont{Gutzwiller}},
  \bibinfo{journal}{Phys. Rev.} \textbf{\bibinfo{volume}{134}},
  \bibinfo{pages}{A923} (\bibinfo{year}{1964}).

\bibitem[{\citenamefont{Gutzwiller}(1965)}]{Gutzwiller3}
\bibinfo{author}{\bibfnamefont{M.~C.} \bibnamefont{Gutzwiller}},
  \bibinfo{journal}{Phys. Rev.} \textbf{\bibinfo{volume}{137}},
  \bibinfo{pages}{A1726} (\bibinfo{year}{1965}).

\bibitem[{\citenamefont{Andrei}(1980)}]{Andrei1}
\bibinfo{author}{\bibfnamefont{N.}~\bibnamefont{Andrei}},
  \bibinfo{journal}{Phys. Rev. Lett.} \textbf{\bibinfo{volume}{45}},
  \bibinfo{pages}{379} (\bibinfo{year}{1980}).

\bibitem[{\citenamefont{Andrei et~al.}(1983)\citenamefont{Andrei, Furuya, and
  Lowenstein}}]{Andrei2}
\bibinfo{author}{\bibfnamefont{N.}~\bibnamefont{Andrei}},
  \bibinfo{author}{\bibfnamefont{K.}~\bibnamefont{Furuya}}, \bibnamefont{and}
  \bibinfo{author}{\bibfnamefont{J.~H.} \bibnamefont{Lowenstein}},
  \bibinfo{journal}{Rev. Mod. Phys.} \textbf{\bibinfo{volume}{55}},
  \bibinfo{pages}{331} (\bibinfo{year}{1983}).

\bibitem[{\citenamefont{Weigmann}(1980)}]{Weigmann}
\bibinfo{author}{\bibfnamefont{P.~B.} \bibnamefont{Weigmann}},
  \bibinfo{journal}{JETP Lett.} \textbf{\bibinfo{volume}{31}},
  \bibinfo{pages}{364} (\bibinfo{year}{1980}).

\bibitem[{\citenamefont{Lanat\`a et~al.}(2015)\citenamefont{Lanat\`a, Yao,
  Wang, Ho, and Kotliar}}]{Our-PRX}
\bibinfo{author}{\bibfnamefont{N.}~\bibnamefont{Lanat\`a}},
  \bibinfo{author}{\bibfnamefont{Y.-X.} \bibnamefont{Yao}},
  \bibinfo{author}{\bibfnamefont{C.-Z.} \bibnamefont{Wang}},
  \bibinfo{author}{\bibfnamefont{K.-M.} \bibnamefont{Ho}}, \bibnamefont{and}
  \bibinfo{author}{\bibfnamefont{G.}~\bibnamefont{Kotliar}},
  \bibinfo{journal}{Phys. Rev. X} \textbf{\bibinfo{volume}{5}},
  \bibinfo{pages}{011008} (\bibinfo{year}{2015}).

\bibitem[{\citenamefont{Lanat\`a}(2010)}]{my-transport-1}
\bibinfo{author}{\bibfnamefont{N.}~\bibnamefont{Lanat\`a}},
  \bibinfo{journal}{Phys. Rev. B} \textbf{\bibinfo{volume}{82}},
  \bibinfo{pages}{195326} (\bibinfo{year}{2010}).

\bibitem[{\citenamefont{Lanat\`a and Strand}(2012)}]{my-transport-2}
\bibinfo{author}{\bibfnamefont{N.}~\bibnamefont{Lanat\`a}} \bibnamefont{and}
  \bibinfo{author}{\bibfnamefont{H.~U.~R.} \bibnamefont{Strand}},
  \bibinfo{journal}{Phys. Rev. B} \textbf{\bibinfo{volume}{86}},
  \bibinfo{pages}{115310} (\bibinfo{year}{2012}).

\bibitem[{\citenamefont{Lanat\`a et~al.}(2008)\citenamefont{Lanat\`a, Barone,
  and Fabrizio}}]{nostro}
\bibinfo{author}{\bibfnamefont{N.}~\bibnamefont{Lanat\`a}},
  \bibinfo{author}{\bibfnamefont{P.}~\bibnamefont{Barone}}, \bibnamefont{and}
  \bibinfo{author}{\bibfnamefont{M.}~\bibnamefont{Fabrizio}},
  \bibinfo{journal}{Phys. Rev. B} \textbf{\bibinfo{volume}{78}},
  \bibinfo{pages}{155127} (\bibinfo{year}{2008}).

\bibitem[{\citenamefont{Lanat{\`a} et~al.}(2015)\citenamefont{Lanat{\`a}, Yao,
  Wang, Ho, and Kotliar}}]{Our-PRB-SB}
\bibinfo{author}{\bibfnamefont{N.}~\bibnamefont{Lanat{\`a}}},
  \bibinfo{author}{\bibfnamefont{Y.-X.} \bibnamefont{Yao}},
  \bibinfo{author}{\bibfnamefont{C.-Z.} \bibnamefont{Wang}},
  \bibinfo{author}{\bibfnamefont{K.-M.} \bibnamefont{Ho}}, \bibnamefont{and}
  \bibinfo{author}{\bibfnamefont{G.}~\bibnamefont{Kotliar}},
  \bibinfo{journal}{Unpublished}  (\bibinfo{year}{2015}).

\bibitem[{\citenamefont{Peschel}(2012)}]{Entanglement-review}
\bibinfo{author}{\bibfnamefont{I.}~\bibnamefont{Peschel}},
  \bibinfo{journal}{Braz. J. Phys.} \textbf{\bibinfo{volume}{42}},
  \bibinfo{pages}{267} (\bibinfo{year}{2012}).

\bibitem[{\citenamefont{Knizia and Chan}(2012)}]{DMET}
\bibinfo{author}{\bibfnamefont{G.}~\bibnamefont{Knizia}} \bibnamefont{and}
  \bibinfo{author}{\bibfnamefont{G.~K.-L.} \bibnamefont{Chan}},
  \bibinfo{journal}{Phys. Rev. Lett.} \textbf{\bibinfo{volume}{109}},
  \bibinfo{pages}{186404} (\bibinfo{year}{2012}).

\bibitem[{\citenamefont{Chou et~al.}(2012{\natexlab{a}})\citenamefont{Chou,
  Pollmann, and Lee}}]{MPPS-RC}
\bibinfo{author}{\bibfnamefont{C.-P.} \bibnamefont{Chou}},
  \bibinfo{author}{\bibfnamefont{F.}~\bibnamefont{Pollmann}}, \bibnamefont{and}
  \bibinfo{author}{\bibfnamefont{T.-K.} \bibnamefont{Lee}},
  \bibinfo{journal}{Phys. Rev. B} \textbf{\bibinfo{volume}{86}},
  \bibinfo{pages}{041105} (\bibinfo{year}{2012}{\natexlab{a}}).

\bibitem[{\citenamefont{Chou et~al.}(2012{\natexlab{b}})\citenamefont{Chou,
  Pollmann, and Lee}}]{MPPS-PRB}
\bibinfo{author}{\bibfnamefont{C.-P.} \bibnamefont{Chou}},
  \bibinfo{author}{\bibfnamefont{F.}~\bibnamefont{Pollmann}}, \bibnamefont{and}
  \bibinfo{author}{\bibfnamefont{T.-K.} \bibnamefont{Lee}},
  \bibinfo{journal}{Phys. Rev. B} \textbf{\bibinfo{volume}{86}},
  \bibinfo{pages}{041105} (\bibinfo{year}{2012}{\natexlab{b}}).

\bibitem[{\citenamefont{Rubtsov et~al.}(2005)\citenamefont{Rubtsov, Savkin, and
  Lichtenstein}}]{ctqmc_hybr-exp_Rubtsov}
\bibinfo{author}{\bibfnamefont{A.~N.} \bibnamefont{Rubtsov}},
  \bibinfo{author}{\bibfnamefont{V.~V.} \bibnamefont{Savkin}},
  \bibnamefont{and} \bibinfo{author}{\bibfnamefont{A.~I.}
  \bibnamefont{Lichtenstein}}, \bibinfo{journal}{Phys. Rev. B}
  \textbf{\bibinfo{volume}{72}}, \bibinfo{pages}{035122}
  (\bibinfo{year}{2005}).

\bibitem[{\citenamefont{Werner et~al.}(2006)\citenamefont{Werner, Comanac, de'
  Medici, Troyer, and Millis}}]{ctqmc}
\bibinfo{author}{\bibfnamefont{P.}~\bibnamefont{Werner}},
  \bibinfo{author}{\bibfnamefont{A.}~\bibnamefont{Comanac}},
  \bibinfo{author}{\bibfnamefont{L.}~\bibnamefont{de' Medici}},
  \bibinfo{author}{\bibfnamefont{M.}~\bibnamefont{Troyer}}, \bibnamefont{and}
  \bibinfo{author}{\bibfnamefont{A.~J.} \bibnamefont{Millis}},
  \bibinfo{journal}{Phys. Rev. Lett.} \textbf{\bibinfo{volume}{97}},
  \bibinfo{pages}{076405} (\bibinfo{year}{2006}).

\bibitem[{\citenamefont{Parcollet et~al.}(2015)\citenamefont{Parcollet,
  Ferrero, Ayral, Hafermann, Krivenko, Messio, and Seth}}]{TRIQS}
\bibinfo{author}{\bibfnamefont{O.}~\bibnamefont{Parcollet}},
  \bibinfo{author}{\bibfnamefont{M.}~\bibnamefont{Ferrero}},
  \bibinfo{author}{\bibfnamefont{T.}~\bibnamefont{Ayral}},
  \bibinfo{author}{\bibfnamefont{H.}~\bibnamefont{Hafermann}},
  \bibinfo{author}{\bibfnamefont{I.}~\bibnamefont{Krivenko}},
  \bibinfo{author}{\bibfnamefont{L.}~\bibnamefont{Messio}}, \bibnamefont{and}
  \bibinfo{author}{\bibfnamefont{P.}~\bibnamefont{Seth}}
  (\bibinfo{year}{2015}), \eprint{cond-mat/1504.01952}.

\bibitem[{\citenamefont{Lanat\`a et~al.}(2015)\citenamefont{Lanat\`a, Deng, and
  Kotliar}}]{Our-Temperature}
\bibinfo{author}{\bibfnamefont{N.}~\bibnamefont{Lanat\`a}},
  \bibinfo{author}{\bibfnamefont{X.-Y.} \bibnamefont{Deng}}, \bibnamefont{and}
  \bibinfo{author}{\bibfnamefont{G.}~\bibnamefont{Kotliar}},
  \bibinfo{journal}{Phys. Rev. B} \textbf{\bibinfo{volume}{92}},
  \bibinfo{pages}{081108} (\bibinfo{year}{2015}).

\bibitem[{\citenamefont{Dirks et~al.}(2013)\citenamefont{Dirks, Schmitt, Han,
  Anders, Werner, and Pruschke}}]{Dirks-comparative-noneq_EPL}
\bibinfo{author}{\bibfnamefont{A.}~\bibnamefont{Dirks}},
  \bibinfo{author}{\bibfnamefont{S.}~\bibnamefont{Schmitt}},
  \bibinfo{author}{\bibfnamefont{J.~E.} \bibnamefont{Han}},
  \bibinfo{author}{\bibfnamefont{F.}~\bibnamefont{Anders}},
  \bibinfo{author}{\bibfnamefont{P.}~\bibnamefont{Werner}}, \bibnamefont{and}
  \bibinfo{author}{\bibfnamefont{T.}~\bibnamefont{Pruschke}},
  \bibinfo{journal}{Europhys. Lett.} \textbf{\bibinfo{volume}{102}},
  \bibinfo{pages}{37011} (\bibinfo{year}{2013}).

\end{thebibliography}

\end{document}